\documentclass[prl,twocolumn,floatfix,showpacs,superscriptaddress]{revtex4}
\usepackage{amsmath}
\usepackage[dvips]{graphicx}

\newcommand{\eq}[2]
{\begin{equation}
    #1
    \label{#2}
  \end{equation}}

\newcommand{\eqnn}[2]
{\begin{equation*}
    #1
    \label{#2}
  \end{equation*}}

\begin{document}

\title{Metal-insulator transitions in the Periodic Anderson Model}
\author{G. Sordi}
\affiliation{Laboratoire de Physique des Solides, CNRS-UMR8502, Universit\'e de Paris-Sud,
Orsay 91405, France.}
\author{A. Amaricci}
\affiliation{Laboratoire de Physique des Solides, CNRS-UMR8502, Universit\'e de Paris-Sud,
Orsay 91405, France.}
\author{M.J. Rozenberg}
\affiliation{Laboratoire de Physique des Solides, CNRS-UMR8502, Universit\'e de Paris-Sud,
Orsay 91405, France.}
\affiliation{Departamento de F\'{\i}sica, FCEN, Universidad de Buenos Aires,
Ciudad Universitaria Pab.I, Buenos Aires (1428), Argentina.}

\date{today}
\begin{abstract}
We solve the Periodic Anderson model in the Mott-Hubbard regime, 
using Dynamical Mean Field Theory. 
Upon electron doping of the Mott insulator, a metal-insulator
transition occurs which is qualitatively similar to 
that of the single band Hubbard model, 
namely with a divergent effective mass and a first 
order character at finite temperatures. 
Surprisingly, upon hole doping, the metal-insulator transition 
is not first order and does not show a divergent mass. 
Thus, the transition scenario of the single band Hubbard model is not generic
for the Periodic Anderson model, even in the Mott-Hubbard regime.
\end{abstract}


\maketitle

The metal-insulator transition in strongly correlated materials remains 
a central problem of modern condensed matter physics \cite{mott,ift}. 
Great progress in its understanding was made possible by the development
of new theoretical approaches such as the 
Dynamical Mean Field Theory \cite{rmp}, 
which is a method that becomes exact in the limit of large lattice connectivity 
\cite{mv}. 
The mean field equations can usually be tackled with a variety of numerical 
approaches which allow to obtain reliable solutions and insights. 
In this context, the Hubbard model, which is probably the simplest model that 
captures a correlation driven metal-insulator transition (MIT), called Mott-Hubbard
transition, 
has received most of the attention in the past 15 years.
As a result of intense investigation, our understanding of the metal-insulator 
transition in that model is now profound.
The studies have unveiled a scenario where, at low temperatures and moderate 
interaction, the half-filled Mott insulator may be driven to a correlated
metallic state through a first order transition\cite{ucs}. 
The transition can occur as a function of correlation strength, temperature
or doping. The first order line ends at finite temperature in a critical point
and the critical region can be described by a Ginzburg-Landau theory \cite{prls}.
This theoretical prediction was experimentally verified in  
experiments on $V_2O_3$ \cite{limelette}.
The Hubbard model is often considered as a minimal model
for the study of rather complicated compounds such as transition metal oxides
and heavy fermion systems.
This is supported by the implicit assumption that the Hubbard model is expected
to be the effective low energy Hamiltonian for a wider class of more realistic
multiband models for strongly correlated electron systems.

On the other hand, a more realistic model
which is 
also widely used in theoretical investigations
of strongly interacting systems, though still schematic, 
is the Periodic 
Anderson model (PAM). 
In the context of correlated electron systems, this model permits to 
describe explicitly both, the localized orbitals, such as the $d$ in 
transition metal oxides or the $f$ in heavy fermion systems, and their 
hybridization to an itinerant electron band
(such as that of $p$ orbitals of oxygen in transition metal oxides). 
In fact, the PAM allows to investigate the various regimes where Mott insulating
states occur, as characterized by the Zaanen-Sawatzky-Allen (ZSA) scheme \cite{zsa}.
They are classified as either Mott-Hubbard insulators
or charge transfer insulators. The first
apply to the early transition metal oxides such as titanates and vanadates, while
the second is relevant for cuprates, such as the high $T_c$ superconductors, and
manganites, which show colossal magnetoresistance \cite{ift}.
In theoretical studies, however, it is often assumed that both
Mott-Hubbard and charge transfer systems may be described at 
low energies by a simpler one band Hubbard model Hamiltonian.

In the present work we shall test the 
putative validity of the Hubbard model as the
effective low energy Hamiltonian of the more realistic Periodic Anderson model.
We shall do this
within a well defined mathematical framework, namely, the  
Dynamical Mean Field Theory (DMFT),
that allows us to obtain essentially exact numerical solutions of the models
(in the statistical Monte Carlo sense).
In particular we shall concentrate on the nature of the (paramagnetic) 
metal-insulator transitions
that occur in the Periodic Anderson model with parameters that set it 
in the Mott-Hubbard regime, and discuss it with respect to the corresponding
scenario that is realized 
in the one band Hubbard model case.
In addition, our results should also be valuable for the interpretation of 
experimental spectroscopies of  strongly correlated transition metal oxides, 
that experienced fantastic improvements in the last decade. In fact, the
analysis of experimental data of systems which have a mixed orbital
character is not always simple when strong correlations are present.
Finally, our work addresses a very relevant issue in regard of the intense effort
that is currently dedicated to the implementation of {\it ab initio} methods
for strongly correlated materials \cite{phystoday} which makes heavy use of the 
DMFT methodology \cite{gabyRMP}.

Among our main results we find that in the case of the electron 
doped driven MIT, the scenario is indeed similar to the one realized in the 
Hubbard model, however, the hole doping scenario 
is qualitatively different. In this case, the correlated metal has 
a resonance peak at the Fermi energy flanked by a Hubbard band, but, 
unlike the Hubbard model scenario, it is not related to the formation of a 
Kondo like resonance and its mass does not diverge at the transition. 
Moreover, and also in contrast to the Hubbard model case, our results indicate 
that this metal-insulator transition is of second order as no signs of
coexistent solutions were observed.
We shall argue that while the metallic state in the former case is a renormalized 
``Brinkman-Rice'' Fermi liquid \cite{br}, the latter can be interpreted 
as liquid of ``Zhang-Rice singlets'' \cite{zr}. 

The Periodic Anderson model Hamiltonian reads,
\eqnn{
\begin{split}
H=&-t\sum_{<ij>\sigma} (p^+_{i\sigma}p_{j\sigma} + p^+_{j\sigma}p_{i\sigma}) + 
\left(\epsilon_p - \mu \right) \sum_{i\sigma} p^+_{i\sigma}p_{i\sigma} \\
 &+(\epsilon_d-\mu)\sum_{i\sigma} d^+_{i\sigma}d_{i\sigma} 
 +t_{pd} \sum_{i\sigma}\left( d^+_{i\sigma}p_{i\sigma} +p^+_{i\sigma}d_{i\sigma} \right) \\
&+U\sum_i  \left(n_{di\uparrow}-\tfrac{1}{2}\right)\left(n_{di\downarrow}-\tfrac{1}{2}\right) 
\end{split}}{PAMHam}
\noindent
where the $d_\sigma$ and $d^+_\sigma$ operators destroy and create electrons
at non-dispersive $d-$orbitals with energy $\epsilon_d$, 
$p_\sigma$ and $p^+_\sigma$ destroy and create electrons at $p-$orbitals
with energy $\epsilon_p$ which form a band with hopping parameter $t$. 
The $p-$ and $d-$orbitals are hybridized with an amplitude $t_{pd}$, 
and the electron correlations are introduced by the Coulomb interaction 
$U$ on the $d-$sites. It is customary to define the charge transfer 
energy $\Delta = \epsilon_d - \epsilon_p$, and $\mu$ is the chemical 
potential. As described in the ZSA scheme this model 
predicts correlated insulating states in two very different regimes: 
at $\Delta << U$ the charge transfer insulator, and at $U \lesssim \Delta$ 
the Mott-Hubbard insulator. The latter is relevant for the early transition 
metal oxides and will be the focus of the present work.

To solve the PAM using DMFT, for simplicity
we adopt a Bethe lattice that corresponds to a semicircular
density of states (DOS) for the $p-$electron band. Setting
the hopping of the $p-$electrons to $t=1/2$, their 
half-bandwidth is equal to one, and fixes the units of the
model.
To set the system in the Mott-Hubbard regime, we adopt 
$\epsilon_d =0$ and $\epsilon_p$ negative, so that the $p-$band 
lies well below the Fermi surface
and is almost full, while the occupation of the local $d-$sites 
will be close to one. The parameter $t_{pd}$ controls the hybridization 
between the orbitals at each lattice site, and permits the delocalization 
of the $d-$electrons. In fact, a finite $t_{pd}$ turns the ``flat'' band 
of $d-$orbitals into a conduction band with mainly $d$ character and 
bandwidth of the order of $t_{pd}^2/\Delta$. 
Now, for a moderate value of the repulsion $U > t_{pd}^2/\Delta$ 
and an occupation of the $d-$site $n_d$ close to one, one expects 
the conduction band to open a correlation gap and the system 
becomes a Mott-Hubbard insulator.

The DMFT equations are most easily derived using the cavity 
method \cite{gks,rmp}, and one obtains the local effective action 
for the $d-$electrons:
\eq{
\begin{split}
{\rm S_{\rm eff}}=&
-\int_0^{\beta}d\tau \int_0^\beta d\tau' \sum_{\sigma} d^+_{\sigma}(\tau)
{\cal G}^{-1}_0(\tau - \tau')d_{\sigma}(\tau') \\
&+U\int_0^{\beta} d\tau \left[n_{d\uparrow}(\tau)-\tfrac{1}{2}\right]
\left[n_{d\downarrow}(\tau) -\tfrac{1}{2}\right] 
\end{split}
}{Seff}
where $d_{\sigma}$ and $d^+_{\sigma}$ correspond to a given (any) site of the lattice.
This equation defines the so called associated impurity problem of the model, that is
subject to a self consistent constraint that reads,
\eqnn{
{\cal G}_0^{-1}(i\omega_n)=
i\omega_n+\mu-\epsilon_d-\frac{{{t^2_{pd}}}}{i\omega_n+\mu-\epsilon_p-t^2G_{pp}}\ .}
{self}
The solution of the quantum impurity problem (\ref{Seff}) gives the local
$d-$electron Green's function $G_{dd}$ and defines a self-energy 
$\Sigma$=${\cal G}^{-1}_0 - G^{-1}_{dd}$.
The local Green's function of the $p-$electrons $G_{pp}$ is 
obtained in terms of $\Sigma$ and the non-interacting 
semicircular DOS $\rho_0$ as:
$$
G_{pp}=\int{ d\epsilon \frac{\rho_0(\epsilon)}{i\omega_n +\mu -\epsilon_p -
\frac{t^2_{pd}}{i\omega_n+\mu-\epsilon_d -\Sigma(i\omega_n)} - \epsilon}}\ .
$$
\noindent

We solve for these equations using two powerful, and in principle
exact, numerical methods: Quantum Monte Carlo (QMC)
and Exact Diagonalization (ED) \cite{rmp}. A similar methodology
was used in the study of a related two-band Hubbard model 
\cite{mutou-ono}.
The QMC is a finite temperature method
and is exact in the statistical sense, while
the ED is at $T=0$ and relies on diagonalization of finite clusters and 
extrapolations to account for the
systematic finite size effects \cite{rmp}.

\begin{figure}
\centering
\includegraphics[width=7cm,clip]{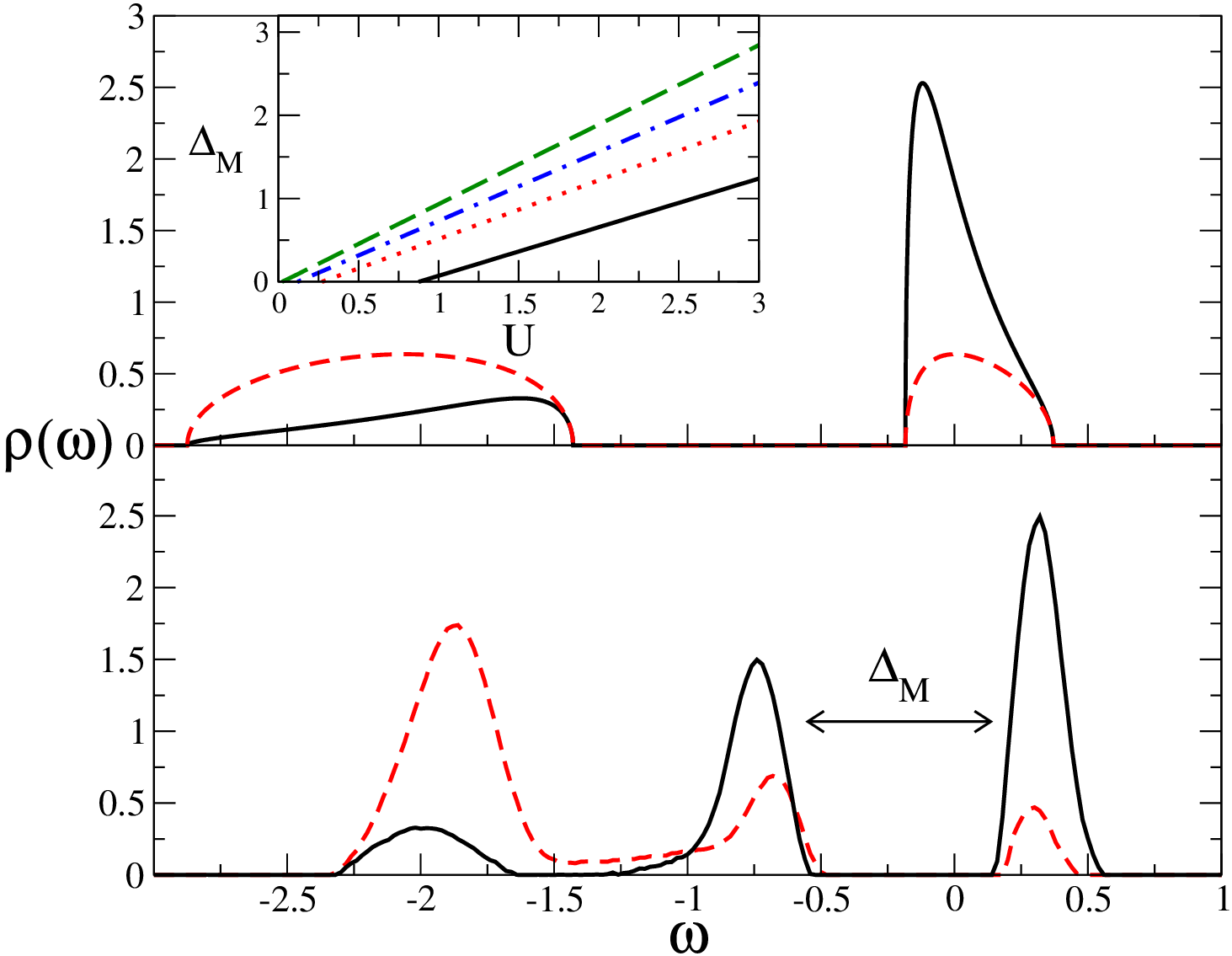}
\caption{
Density of states for the $p-$ and $d-$electrons (dashed and solid line) 
for $\Delta=1$, $t_{pd}=0.9$. 
Upper panel shows the $U=0$ case with $\mu=0.529$ that gives a total occupation
of three particles below the Fermi energy. 
The lower panel shows analytically continued QMC data for the Mott insulator
state with $U=2$, $T=1/64$ and $\mu=1.079$ that
gives $n_{tot}=3$.  
The double arrow head line indicates the large Mott gap $\Delta_{M}$.
The inset shows $\Delta_M(U)$ for different positions of the
of the $p-$electron band $\epsilon_p=-6,-3,-2,-1$ 
(top to bottom). The results are obtained with ED of finite clusters of $N_s$ sites
and the data shown is for the limit $N_s \rightarrow\infty$.
}
\label{fig1}
\end{figure}
We shall begin our discussion with the behavior of the 
density of states
of the model in different regimes.
For $U=0$ and at $t_{pd}=0$ the system is an insulator since
neither the $d-$orbitals at the Fermi energy can conduct (because
they are localized), nor the $p-$band can conduct (because it is
full and well beneath the Fermi surface).
At a finite hybridization $t_{pd}$, however, the 
system becomes metallic, as the $p$ and $d$ orbitals form
a partially filled band at the Fermi energy with mixed $p$ and $d$ character. 
In Fig.\ref{fig1} we show the comparison of the $p-$ and $d-$electron DOS.
The one carrying most of the
spectral intensity at low frequencies is the $d-$electron DOS $\rho_d(\omega)$,
since the bare atomic energy of the $d-$orbitals is at
the Fermi energy. 
The lower panel of the figure shows the dramatic effect of 
correlations; when the interaction $U$ is increased, 
a rather large gap opens in the DOS at the Fermi energy,
driving the system to a Mott insulator state.
The gap is of order $U$ and results from the
high energetic cost of double occupation of the $d$ site.
An interesting effect is that the size of the Mott gap $\Delta_M$
may be substantially renormalized. In the inset of Fig.\ref{fig1}
we show the variation of $\Delta_M(U)$ upon increasing the distance
of the $p-$band with respect to the $d-$electron energy.
Notice that the gap $\Delta_M(U)$ is always smaller than the bare $U$, 
and become equal only asymptotically when $\epsilon_p \to -\infty$.
This renormalization effect is of relevance to the difficult problem of the 
determination of the effective value of $U$ in realistic
{\it ab initio} calculations using DMFT \cite{phystoday}.

\begin{figure}
\centering
\includegraphics[width=7cm,clip]{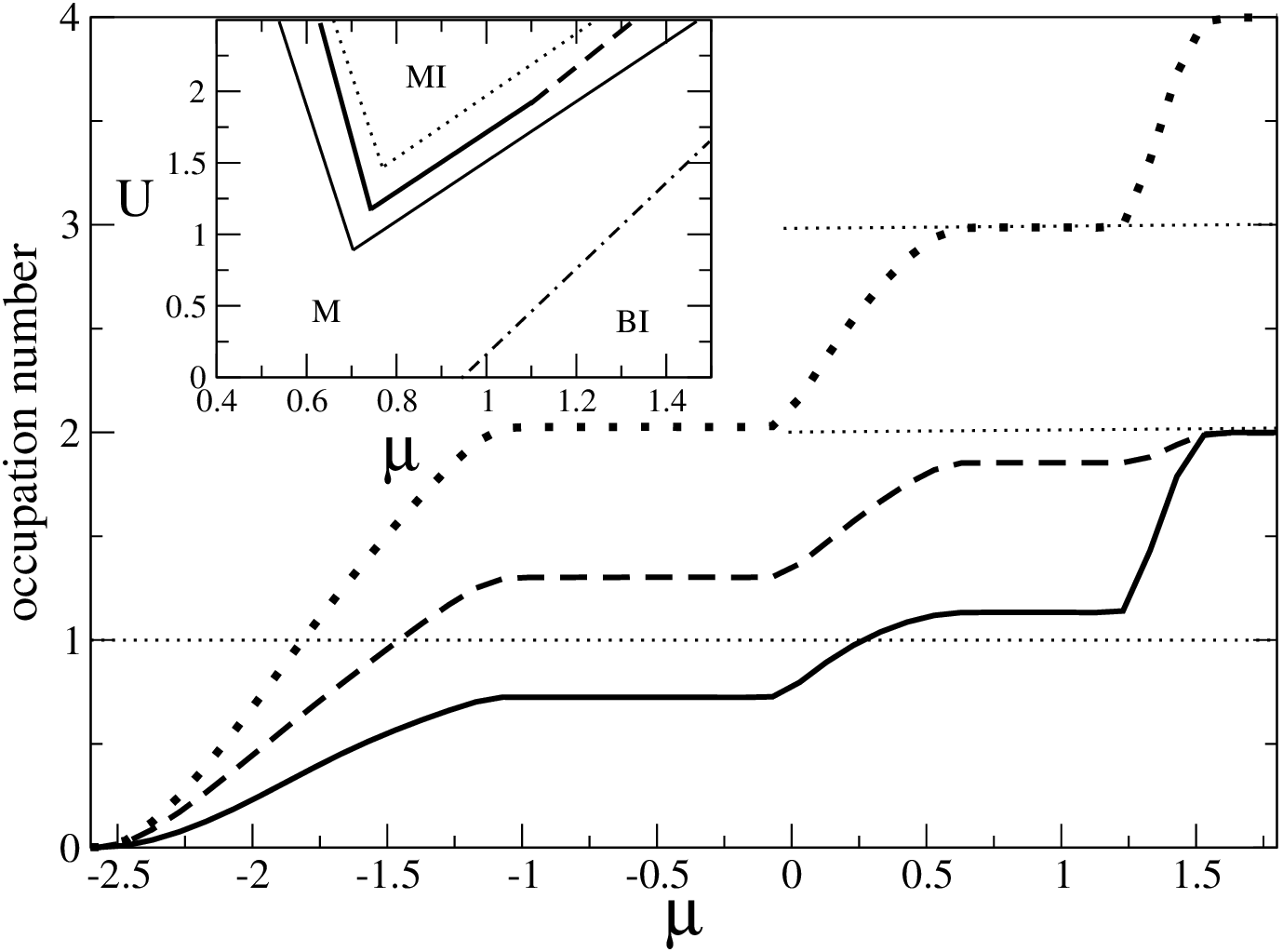}
\caption{$n_d$ (solid line), $n_p$ (dashed) and $n_{tot}$ 
(dotted) as a function of $\mu$, for $U=2$.
The data are from QMC at $\Delta=1$, $t_{pd}=0.9$ and $T=1/64$.
The plateaux at $n_{tot}$=2 and 4 are band insulator (BI) states.
The one at $n_{tot}$=3 is the Mott insulator (MI).
The inset shows the phase diagram in the $U$-$\mu$ plane. 
The boundary lines are for $T$=1/20 (dotted), $T$=1/64 (thick solid) 
and $T=0$ (thin solid).
The dashed thick line segment at $T=1/64$ denotes the region of the MIT boundary
where the QMC data show coexistence of a metal and an insulating solution.
The $T$=0 data are from extrapolated ED calculations. This method is not 
suited for the study of coexistence of solutions at $T=0$.  
The dash-dot line denotes the transition from a metal (M) to the band insulator
(BI) at $n_{tot}=4$.
}
\label{fig2}
\end{figure}

The Mott insulator can be destabilized by either
particle or hole doping.  
Therefore the system has two doping driven
metal-insulator transitions. In the one band Hubbard model,
the two transitions have the same character, however, as we shall
see, this is not the case in the present model.
In Fig.\ref{fig2} we
show the occupation $n_d$ and $n_p$ of the $d$ and $p$ sites
for $U=2$.
The plateaux that appear around $0.7\lesssim \mu \lesssim 1.2$ 
indicate the onset of the 
incompressible Mott insulating state when correlations are strong.
While the Mott insulator is associated with the energy
cost of doubly occupying the local $d-$orbital, it is interesting 
to notice that the Mott plateau does not occur 
exactly at $n_d=1$ but at a higher value, which depends on the hybridization. 
The Mott state is in fact found when the total number of
particles per unit cell is exactly equal to three. 
Thus, the object that becomes localized due to the strong correlations
is not simply a $d-$electron, but a composite object with mixed $d$ and
$p$ character.
The inset of the figure shows the phase diagram in the $U$-$\mu$ plane,
that maps the region of the Mott insulator phase and the transition
lines to correlated metallic states.

\begin{figure}
\centering
\includegraphics[width=8cm,clip]{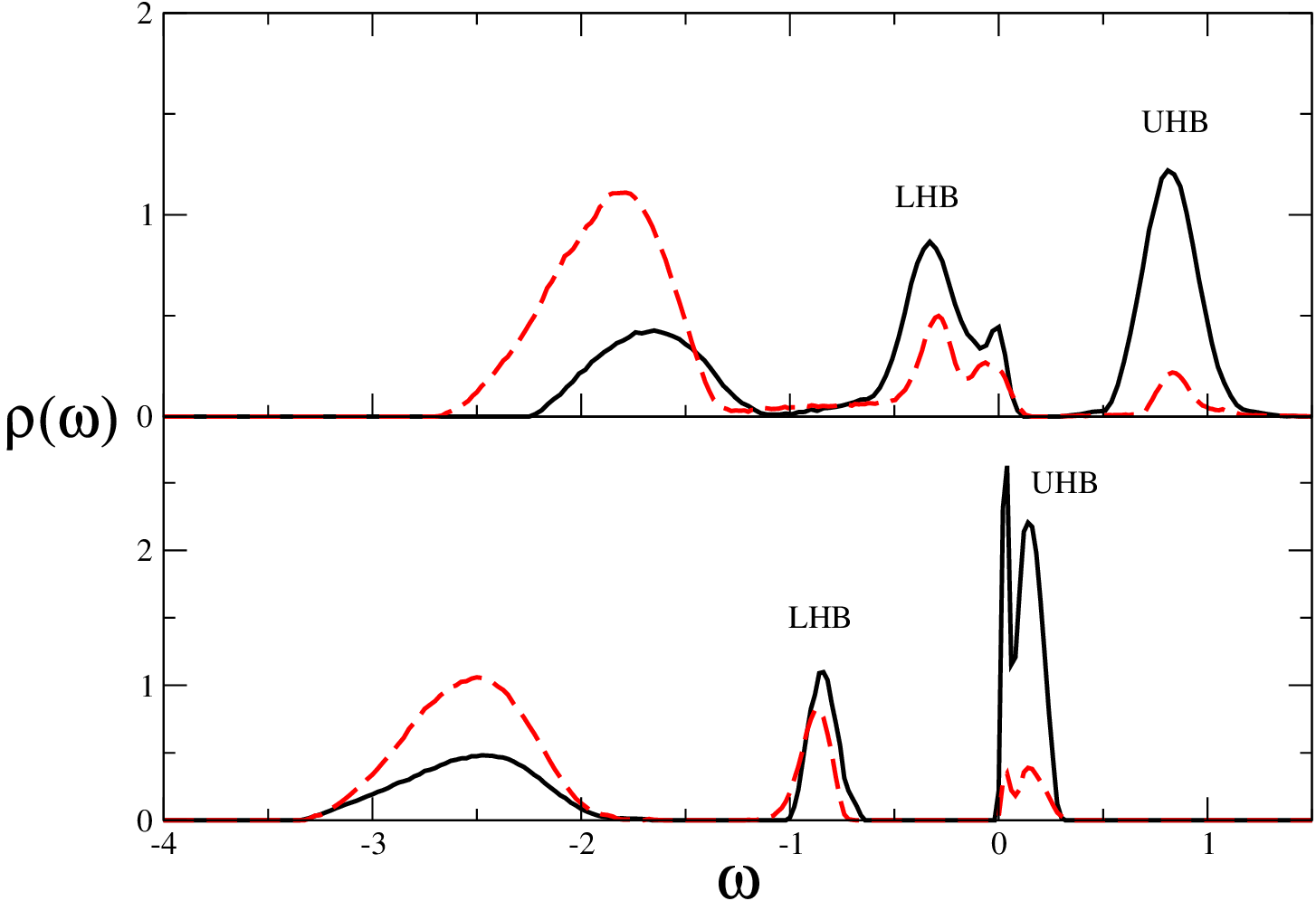}
\caption{Density of states of $p-$ and $d-$electrons (dashed and solid line) 
for $U=2$, $\Delta=1$, $t_{pd}=0.9$ and $T=1/64$, as obtained from QMC.  
{\it Upper panel}: $\mu=0.554$, which corresponds to tiny hole doping.
{\it Lower panel}: $\mu=1.234$, which corresponds to tiny particle doping.}
\label{fig3}
\end{figure}

In Fig.\ref{fig3} we show the DOS for the $p$ and $d$ electrons in the
metallic states that are obtained by either particle or hole doping of
the Mott insulator. In both cases one finds that the
DOS clearly show the emergence of a correlated small
quasiparticle peak at the Fermi energy.
The occurrence of a narrow quasiparticle peak at the Fermi energy that
is flanked by 
large Hubbard bands which are separated by an energy of order $U$ 
is a hallmark
result of the solution of the Hubbard model within DMFT \cite{rmp}. 
Thus, one may be led to conclude that the MIT in the PAM shares the same 
qualitative features.
Rather surprisingly, this expectation is only fully confirmed for the case
of particle doping, but the MIT scenario in the hole doped case is
strikingly different.
Upon particle doping of the Mott insulating state we have confirmed
that there is a small region of parameters  at the MIT boundary
where two coexistent solutions,
one metallic and one insulating, are found. 
In addition, the numerical solutions show critical
slowing down of the convergence speed of the 
self-consistency close to the transition.
These two features were also observed in the previous studies of
the finite $T$ first order MIT in the Hubbard model \cite{prls}.
In contrast, in the hole doped case ($n_{tot} < 3$) we found no trace of
coexistent solutions down to $T=1/128$. 
The solution seems to be always unique, which implies
that the transition is of second order, i.e., through a quantum critical line
in the $U$-$\mu$ plane.

\begin{figure}
\centering
\includegraphics[width=8cm,clip]{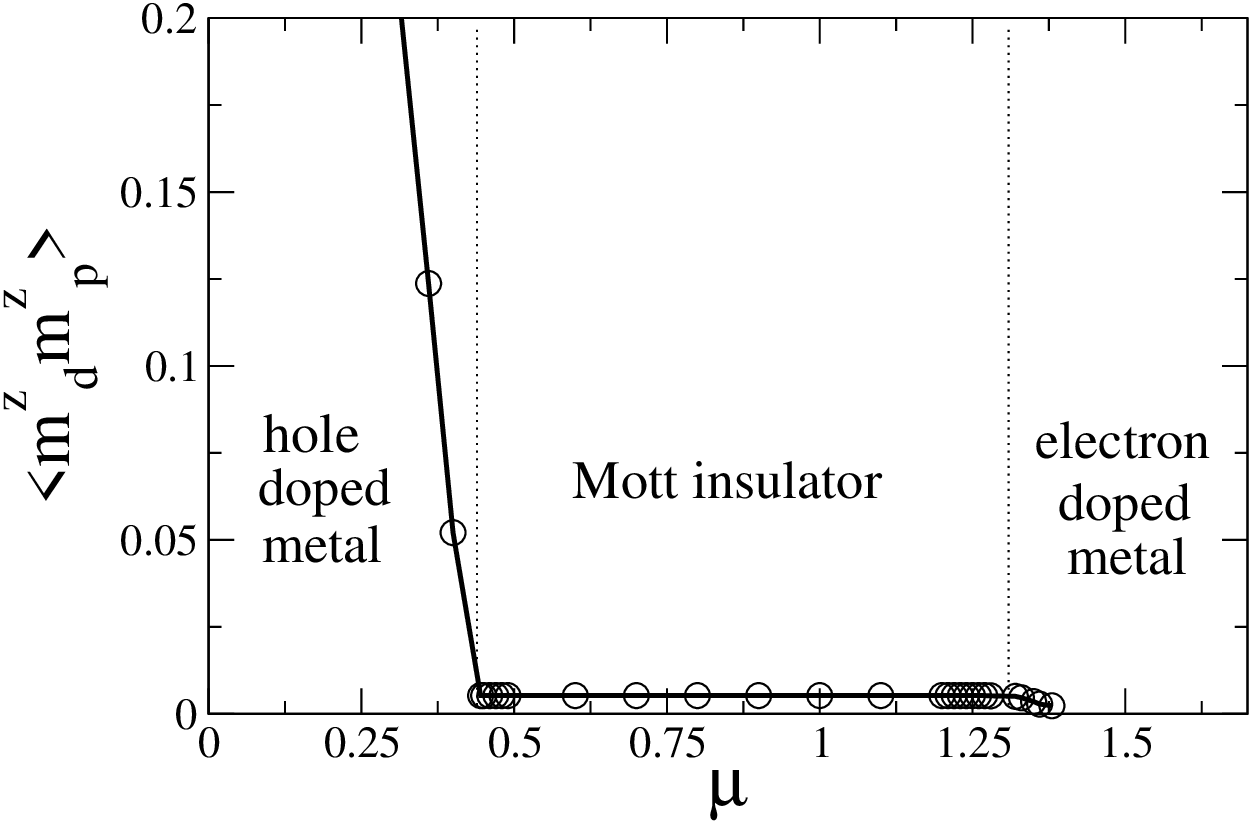}
\caption{$\langle m^z_d m^z_p \rangle$ as a function of $\mu$, 
for $U=2$, $\Delta=1$, $t_{pd}=0.9$. 
The results are obtained from ED with a cluster of 16 sites.}
\label{fig4}
\end{figure}

Insight on the physical reason for the qualitative difference observed upon
particle or hole doping is obtained from the behavior of the observables
that measures the $d$-$p$ correlations. In Fig.\ref{fig4}
we show the behavior of magnetic correlation between the $d$ and
$p$ electrons 
$\langle m^z_d m^z_p \rangle = \langle 
(n_{d\uparrow} - n_{d\downarrow})
(n_{p\uparrow} - n_{p\downarrow}) \rangle $ 
across the metal-insulator transitions.
As the results show, the value of $\langle m^z_d m^z_p \rangle$ is 
sizable on the hole doped metal, while becomes 
negligible on the particle doped
side. 
In the particle doped case $\langle m^z_d m^z_p \rangle$ is
negligible because the $p-$band is already full and the extra particles 
mainly go to occupy $d-$sites. 
Thus, the $p-$sites do not get an induced magnetic
moment and they cannot screen the magnetic moment of the local $d-$sites. Thus
the magnetic correlations develop directly among neighboring $d$ orbitals.
These correlations are of antiferromagnetic character due to the superexchange
mechanism and are analogous to those created between neighboring sites in the
Hubbard model case. 
Thus we can understand that for particle doping the character
of the MIT in the Periodic Anderson and Hubbard models is in fact the same;
the $p-$sites merely allow the charge fluctuation (and thus the delocalization)
of the $d-$electrons, but they do not couple magnetically and do not screen the 
local moments.

In striking contrast, upon hole doping the scenario is quite different.
In this case, the system finds it is energetically
favorable to create holes in the $p-$band and magnetically bound the holes to
the local moment of the correlated $d-$site. This feature is 
reminiscent of
the ``Zhang-Rice'' singlet formation \cite{zr} and leads to the emergence of a 
quasiparticle peak at the Fermi energy as the singlets delocalize.

In conclusion, we have investigated the doping driven metal-insulator transition
in the Periodic Anderson model in the Mott-Hubbard regime. 
We found that the size of the Mott gap can be significantly 
renormalized by hybridization effects. In addition, we found that while
both correlated metallic states at small doping show a 
small quasiparticle peak at the Fermi energy, the
nature of the MIT is qualitatively different on each side. 
In the particle doped side the quasiparticle peak is associated with a 
Kondo-like resonance and the MIT shares the same qualitative nature 
of the first order transition found in the one band Hubbard model.
In contrast, on the hole doped side, the quasiparticle peak 
is associated with the formation of ``Zhang-Rice'' singlets
and the transition is second order.
Thus, our study demonstrates that, even in relatively simple situations,
the one band Hubbard model should not be automatically considered
the low energy effective model of more complicated
multi-orbital systems. The investigation of the physical nature of the
``Zhang-Rice'' correlated metal is a very interesting problem open
for future investigations.

We acknowledge M. Gabay for useful discussions.


%
\end{document}